\documentclass[pra,letterpaper,aps,10pt,superscriptaddress,twocolumn,floatfix,showpacs]{revtex4-1}
\usepackage{amsmath,subfig,graphicx,amssymb,braket,xcolor,upgreek}
\usepackage[colorlinks, linkcolor=blue, citecolor=blue, urlcolor=blue, breaklinks=true]{hyperref}
\usepackage{microtype}
\usepackage[english]{babel}
\usepackage{float}

\begin{document}

\title{Subradiance in Multiply Excited States of Dipole-Coupled V-Type Atoms}
\author{Raphael Holzinger}
\affiliation{Institut f\"ur Theoretische Physik, Universit\"at Innsbruck, Technikerstr. 21a, A-6020 Innsbruck, Austria}
\author{Laurin Ostermann }
\affiliation{Institut f\"ur Theoretische Physik, Universit\"at Innsbruck, Technikerstr. 21a, A-6020 Innsbruck, Austria}
\author{Helmut Ritsch}
\affiliation{Institut f\"ur Theoretische Physik, Universit\"at Innsbruck, Technikerstr. 21a, A-6020 Innsbruck, Austria}
\date{\today}

\begin{abstract}
We generalize the theoretical modeling of collective atomic super- and subradiance to the multilevel case including spontaneous emission from several excited states towards a common ground state. We show that in a closely packed ensemble of $N$ atoms with $N-1$ distinct excited states each, one can find a new class of non-radiating dark states,, which allows for long-term storage of $N-1$ photonic excitations. Via dipole-dipole coupling only a single atom in the ground state is sufficient in order to suppress the decay of all $N-1$ other atoms. By means of some generic geometric configurations, like a triangle of V-type atoms or a chain of atoms with a $J=0 \to J=1$ transition, we study such subradiance including dipole-dipole interactions and show that even at finite distances long lifetimes can be observed. While generally hard to prepare deterministically, we identify various possibilities for a probabilistic preparation via a phase controlled laser pump and decay.
\end{abstract}

\maketitle

\section{Introduction}
Quantum fluctuations in the electromagnetic vacuum field inevitably lead to energy dissipation from excited atomic states via the spontaneous emergence of photons~\cite{dirac1927quantum} known as spontaneous emission. In a quantum electrodynamics  treatment the probability for this process and its corresponding decay rate $\Gamma = \omega_0^3 \mu^2 / (3 \pi \epsilon_0 \hbar c^3)$  was first derived by Weiskopf and Wigner~\cite{weisskopf1935probleme}. It is proportional to the third power of the transition energy between the excited and lower lying state as well as to the square of the transition dipole moment between those two states.

As there is only one electromagnetic vacuum, atoms in close proximity will experience correlated fluctuations inducing cooperative effects in their dissipative behavior. By means of constructive as well as destructive interference of the emerging photons the collective spontaneous emission rates are drastically modified as a function of distance~\cite{lehmberg1970radiation,lehmberg1970radiation2,ficek1987quantum,agarwal2001vacuum}. A strongly increased spontaneous emission is dubbed 'superradiance' while a decreased rate is referred to as 'subradiance'~\cite{gross1982superradiance}.

Due to the quantum nature of atomic excitations, they can be delocalized and distributed over an entire atomic ensemble, exhibiting highly multi-partite entanglement~\cite{lukin2000entanglement,chou2005measurement,plankensteiner2015selective}. Well known examples are the single-excitation Bell states of two atoms~\cite{cabrillo1999creation,raimond2001manipulating}, the W-state~\cite{eibl2004experimental,zou2002generation} and many others.

\begin{figure}[th]
	\centering
 	\includegraphics[width=0.95\columnwidth]{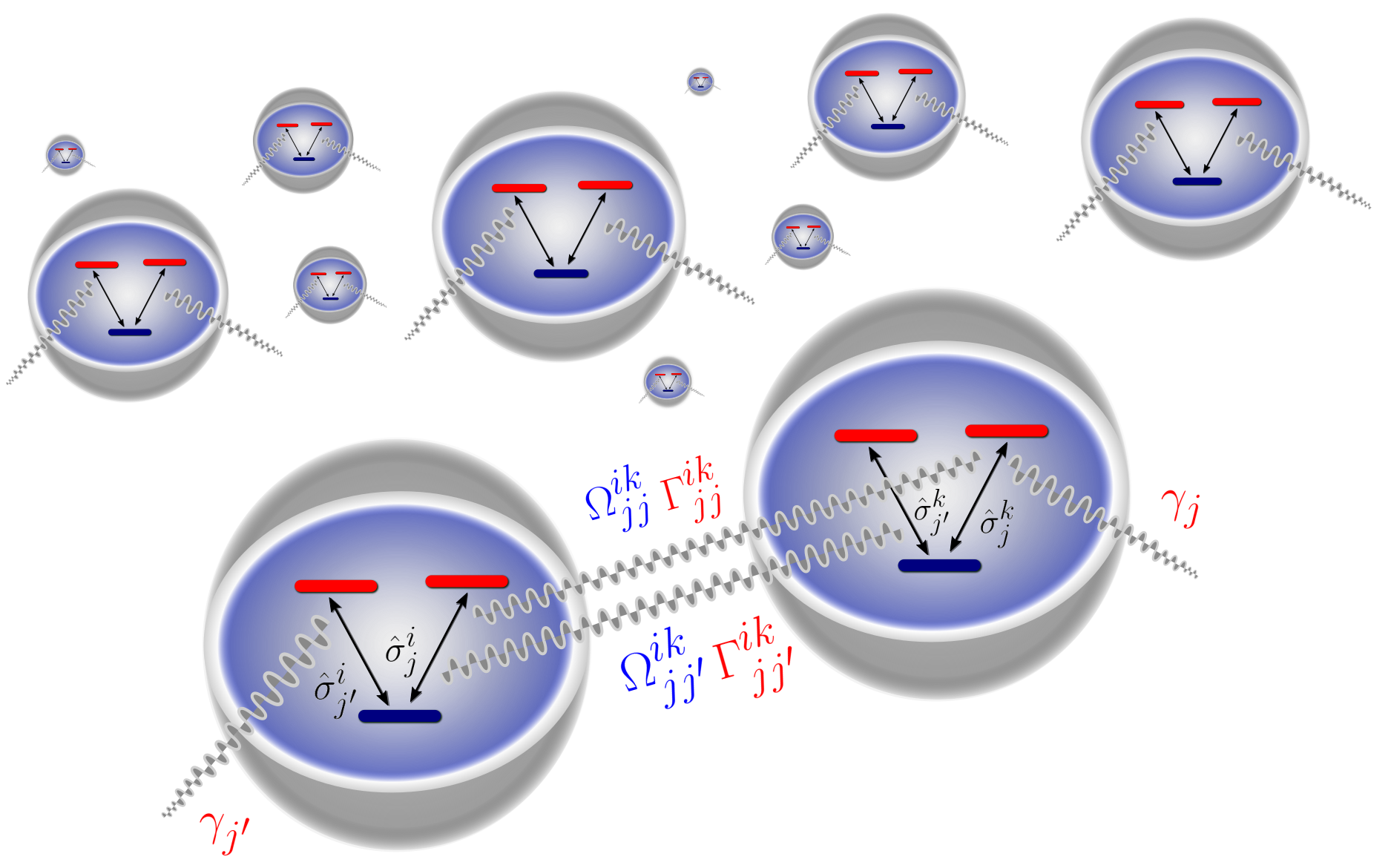}
	\caption{{\textit{Model.} We consider a collection of $N$ identical multilevel atoms separated by a finite distance, which are coupled to the quantized electromagnetic vacuum field. Each atom $i$ features $N-1$ excited states $\ket{e_j}$ with independent transitions to a common ground state $\ket{g}$, represented by ${\sigma}_j^i$. The collective decay rates are given by $\Gamma_{jj'}^{ik}$ whereas the collective energy shifts are written as $\Omega_{jj'}^{ik}$. The individual spontaneous emission rate for transition $j$ in all atoms is $\gamma_j$.}}
	\label{model}
\end{figure}
Depending on the geometry of the atomic ensemble as well as on the local phase difference of the excitation amplitudes between the atoms, such delocalized  excitation states can feature either super- or subradiance. For instance, for two closely spaced atoms ($d \ll \lambda_0 = 2 \pi c/\omega_0$), the symmetric Bell state $\ket{+} = \left( \ket{eg} + \ket{ge} \right)/\sqrt{2}$ is superradiant, while its asymmetric analogue $\ket{-} = \left( \ket{eg} - \ket{ge} \right) / \sqrt{2}$ is strongly subradiant and decouples from the radiation field completely at distances close to zero~\cite{dicke1954coherence}. This leads to the term 'dark state'. Because of the fact that their lifetime is often orders of magnitude longer than typical experimental cycles, those dark states are a valuable resource in quantum information storage and processing~\cite{fleischhauer2002quantum, chaneliere2005storage}.

While subradiant states of dense atomic ensembles are easy to identify theoretically~\cite{temnov2005superradiance,asenjo2017exponential}, they have been quite elusive and hard to find in concrete experiments~\cite{guerin2016subradiance,bromley2016collective}, with directional emission patterns as one of the signatures of destructive interference leading to subradiance~\cite{bhatti2018directional}. Besides the influence of motion and various dephasing mechanisms, it was recently pointed out, that the complex level structure of typical atoms beyond a two-level approximation will often prevent the appearance of perfectly dark states~\cite{hebenstreit2017subradiance}.  In particular, for excited atomic states, which can decay to different lower states via more than one decay channel, the observation of subradiance is much more challenging. It can be easily shown that for a system of two $\Lambda$-type atoms no dark state can be found, as both decay channels need to be blocked via interference, which cannot be achieved simultaneously. However, in earlier work~\cite{hebenstreit2017subradiance} we could show that an ensemble of $N$ $N$-level atoms with and $N-1$ independent decay channels from the excited state to $N-1$ different ground states, a unique perfectly dark state, can be identified. This completely anti-symmetric dark state has remarkable entanglement and symmetry properties making it a promising candidate for quantum information applications.

In this paper, we investigate a related system, namely the inverted energy level configuration involving $N$ atoms where one ground state is coupled to $N-1$ excited states $\ket{i} = \ket{s_i} = \ket{e_i}$. Each upper state can decay independently to a common ground state $\ket{0} = \ket{s_0} = \ket{g}$. Again the totally anti-symmetric state is a dark state of a similar form
\begin{equation} \label{dark1}
| \psi_{\mathrm{d}}^{N} \rangle=\frac{1}{\sqrt{N !}} \sum_{\pi \in S_{N}} \operatorname{sgn}(\pi) \bigotimes_{i} | s_{\pi(i)} \rangle.
\end{equation}
Here the sum runs over all permutations $\pi$ of $N$ elements. Using a spatially symmetric configuration of three atoms we will show below, that this $N$-level state of $N$ atoms is subradiant as well as an eigentstate of the Hamiltonian.

\section{Model}
Let us consider a collection of $N$ identical V-level type atoms at fixed positions $\left \lbrace \vec r_i \right \rbrace_{i=1}^N$. Each atom features $N-1$ excited states $\left \lbrace \ket{e_j} \right \rbrace_{j=1}^{N-1}$ at energies $\omega_j$ with dipole coupling to a common ground state $\ket{g}$ via a transition dipole moment of $\vec \mu_j$.

The combined Hamiltonian of the atoms and the electromagnetic field is given by
\begin{equation} \label{Hamiltonian1}
	H = H_\mathrm{A} + H_\mathrm{F} + H_\mathrm{int}
\end{equation}
with the atomic part $H_\mathrm{A} = \sum _ { i = 1 } ^ { N } \sum _ { j = 1 } ^ { N - 1 } \omega_j \sigma _ { j } ^ { i + } \sigma _ { j} ^ { i - }$ and the field $ H_\mathrm{F} = \sum _ { \vec { k } , \lambda } \omega _ { k } {a} _ { \vec { k } , \lambda } ^ { \dagger } a _ { \vec { k } , \lambda }$.

The interaction between the atoms and the field in dipole approximation is then
\begin{equation}
H_\mathrm{int} = - \sum _ { i = 1 } ^ { N } \sum _ { j = 1 } ^ { N - 1 } \left( \vec { \mu _ { j } ^ { i } } \sigma _ { j } ^ { i + } \cdot {\vec { E }} \left( \vec { r } _ { i } \right) + \mathrm{h.c.} \right),
\end{equation}
where $\vec{E} \left( \vec{r}_i \right)$ is the quantized electromagnetic field. When particularizing to $N=3$ below, we will consider a situation where the transition dipole matrix elements inside each atom are mutually orthogonal and real, that is
\begin{equation}
	\mu _ { j } ^ {  { i } } \cdot \mu _ { j ^ { \prime } } ^ {  { i } } = 0.
\end{equation}

After tracing out the electromagnetic field modes in a standard quantum optics fashion assuming the field in its vacuum state~\cite{gardiner2004quantum,ficek2002entangled,agarwal2001vacuum,moy1999born} the system dynamics can be described by the master equation
\begin{equation} \label{master}
\dot \rho = i \left[ \rho, H \right] + \mathcal { L } \left[ \rho \right]
\end{equation}
with the effective Hamiltonian including dipole-dipole interaction
\begin{equation} \label{Hamiltonian}
	H = \sum _ { i = 1 } ^ { N } \sum _ { j = 1 } ^ { N -1 } \omega _ { j }   \sigma  _ { j } ^ { i + } \sigma _ { j } ^ { i - } + \sum _ { i \neq k } ^ { N } \sum _ { j , j ^ { \prime } } ^ { N - 1 } \Omega _ { j j ^ { \prime } } ^ { i k }   \sigma _ { j } ^ { i + }  \sigma _ { j ^ { \prime } } ^ { k - }
\end{equation}
and the Liouvillian in Lindblad form
\begin{equation} \label{Liouvillian}
\mathcal { L } \left[ \rho \right] = \sum _ { i , k } ^ { N } \sum _ { j , j ^ { \prime } } ^ { N - 1 } \Gamma _ { j j ^ { \prime } } ^ { i k } \left( 2 \sigma _ { j } ^ { i - }   \rho \sigma  _ { j ^ { \prime } } ^ { k + }  -  \sigma _ { j  }  ^ { i + }   \sigma _ { j ^\prime} ^ { k - }  \rho -  \rho \sigma _ { j } ^ { i + }   \sigma _ { j^\prime } ^ { k - } \right),
\end{equation}
where $\sigma^{i \pm}_j$ denotes the rising (lowering) operator of the $j$-th transition in the $i$-th atom.

The coherent part of the dipole-dipole interaction induces energy shifts (see Fig. \ref{model}) due to the couplings
\begin{multline} \label{coherent}
	\Omega _ { j j ^ {\prime } } ^ { i k } = \frac { 3 \sqrt { \gamma _ { j } \gamma _ { j ^ { \prime } } } } { 2 } \left[ \left( \vec { \mu } _ { j } ^ { i } \cdot \vec { \mu } _ { j ^ { \prime } } ^ { k } \right) P _ { R } \left( k _ { 0 } r _ { i k } \right) \right. \\
	\left. - \left( \vec { \mu } _ { j } ^ { i } \cdot \vec { r } _ { i k } \right) \left( \vec { \mu } _ { j ^ { \prime } } ^ { k } \cdot \vec { r } _ { i k } \right) Q _ { R } \left( k _ { 0 } r _ { i k } \right) \right],
\end{multline}
while the incoherent collective dissipation is characterized by
\begin{multline} \label{dissipative}
	\Gamma _ { j j ^ { \prime } } ^ { i k } = \frac { 3 \sqrt { \gamma _ { j } \gamma _ { j ^ { \prime } } } } { 2 } \left[ \left( \hat { \mu } _ { j } ^ { i } \cdot \hat { \mu } _ { j ^ { \prime } } ^ { k } \right) P _ { I } \left( k _ { 0 } r _ { i k } \right) \right. \\
	\left. - \left( \hat { \mu } _ { j } ^ { i } \cdot \hat { r } _ { i k } \right) \left( \hat { \mu } _ { j ^ { \prime } } ^ { k } \cdot \hat { r } _ { i k } \right) Q _ { I } \left( k _ { 0 } r _ { i k } \right) \right].
\end{multline}

\begin{figure}
	 \centering
	 \subfloat{\includegraphics[width=.25\textwidth]{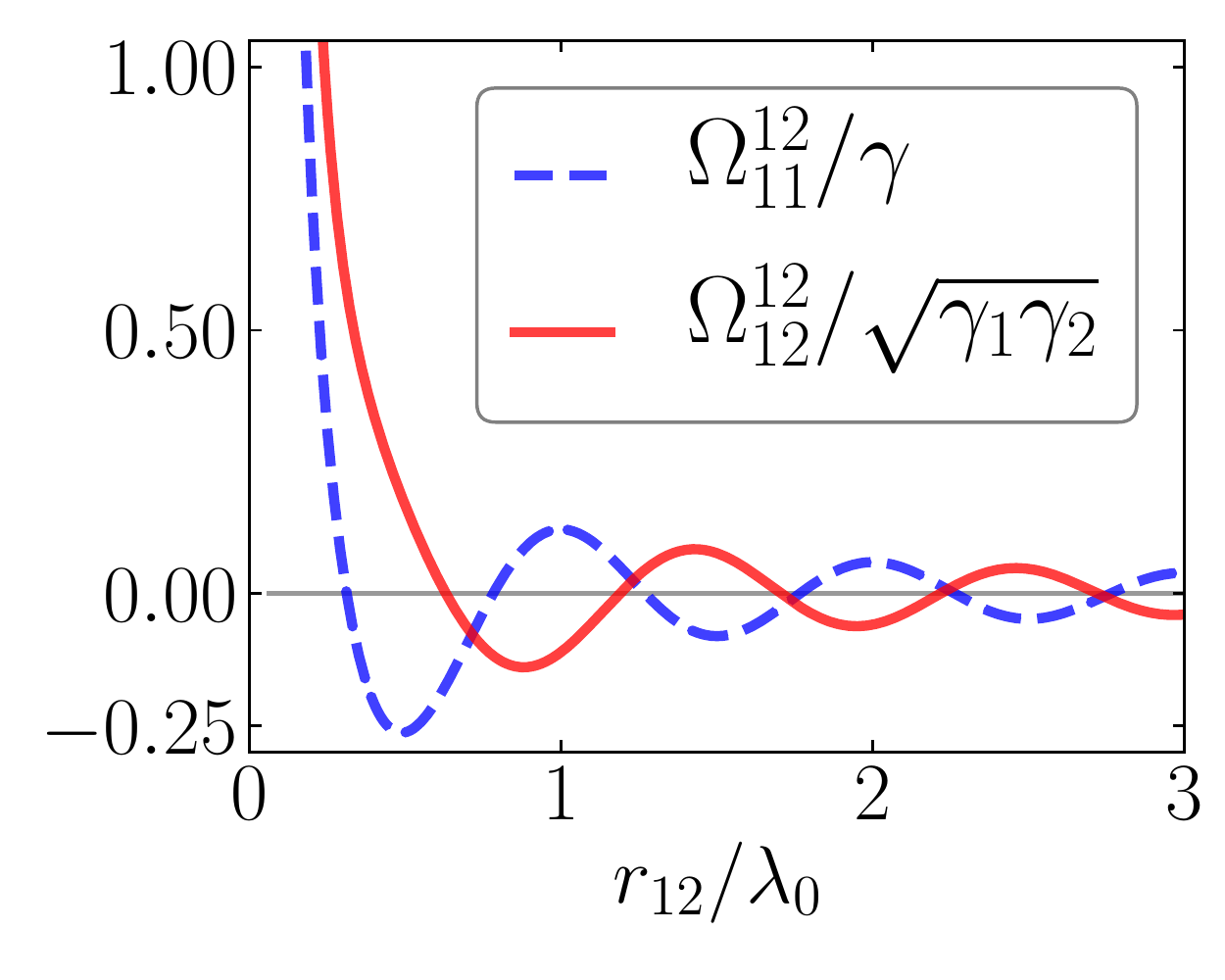}}
	 \subfloat{\includegraphics[width=.25\textwidth]{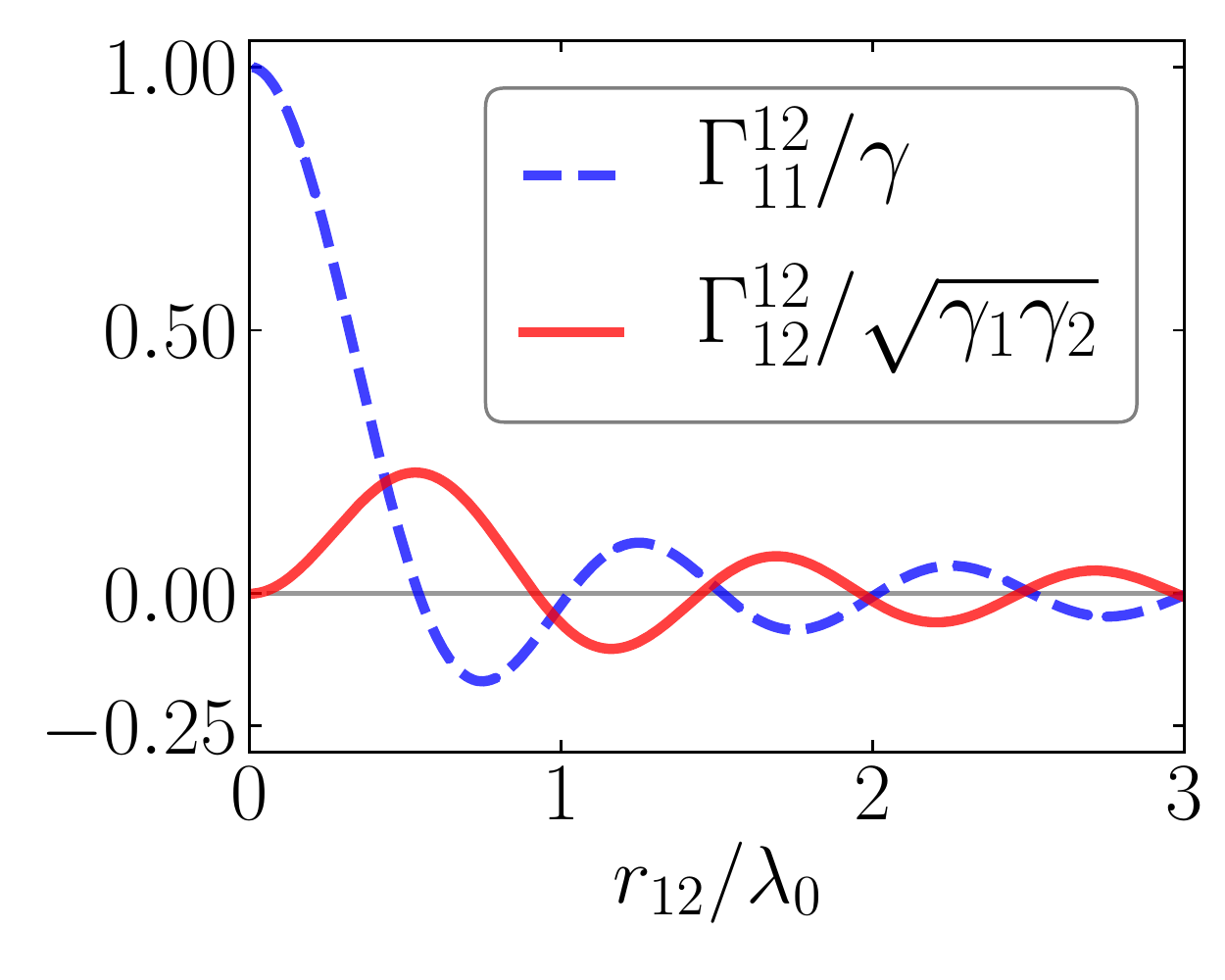} }
	 \caption{\textit{Collective Couplings.} Coherent and dissipative dipole-dipole coupling coefficients as a function of the interatomic distance with $\theta_1 = \pi/4$, $\theta_2 = 3\pi/4$ for $\phi = 0$ (see Fig. \ref{geometry}). The blue dashed lines represent the coupling of two neighbouring and parallel dipole moments whereas the red lines represent couplings of orthogonal dipole moments which appear due to their indirect interaction via the same vacuum field.}
	 \label{couplings}
\end{figure}

\begin{figure}[th]
	\centering
 	\includegraphics[width=0.95\columnwidth]{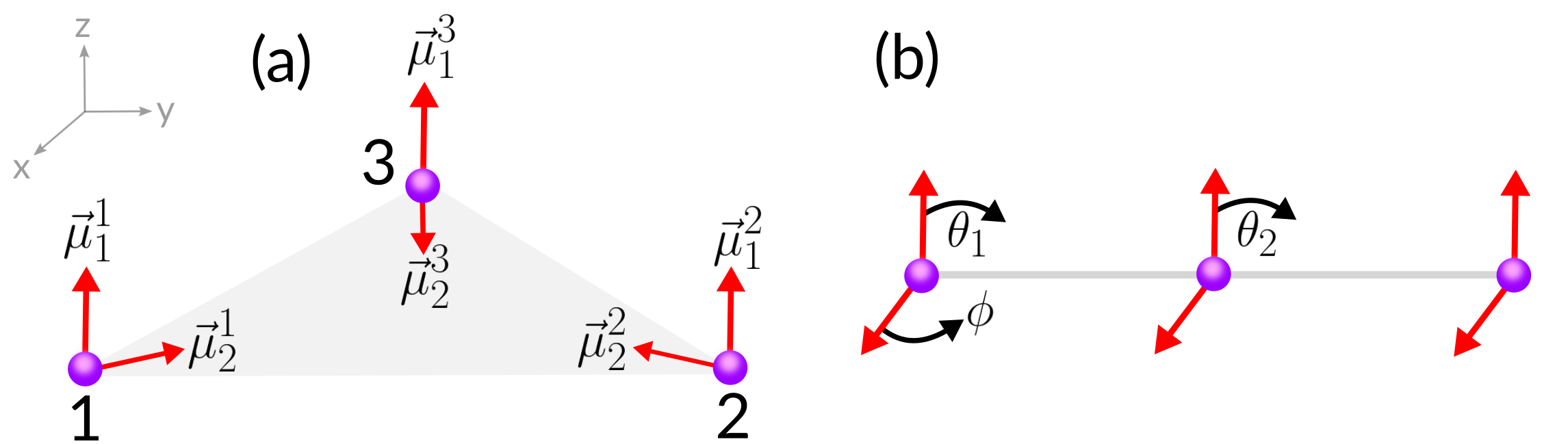}
	\caption{{\textit{Geometries.} We consider (a) an equilateral triangle and (b) a linear chain of three 3-level atoms. Here $\vec{\mu}^i_j$ represents the dipole orientation of the $j$-th transition in atom $i$.}}
	\label{geometry}
\end{figure}

Furthermore,  for brevity we have introduced the functions
\begin{eqnarray}
	P _ { R } ( \xi ) =& \frac { \cos \xi } { \xi } - \frac { \sin \xi } { \xi ^ { 2 } } - \frac { \cos \xi } { \xi ^ { 3 } }, \\
	P _ { I } ( \xi ) =& \frac { \sin \xi } { \xi } + \frac { \cos \xi } { \xi ^ { 2 } } - \frac { \sin \xi } { \xi ^ { 3 } }, \\
	Q _ { R } ( \xi ) =& \frac { \cos \xi } { \xi } - 3 \frac { \sin \xi } { \xi ^ { 2 } } - 3 \frac { \cos \xi } { \xi ^ { 3 }}, \\
	Q _ { I } ( \xi ) =& \frac { \sin \xi } { \xi } + 3 \frac { \cos \xi } { \xi ^ { 2 } } - 3 \frac { \sin \xi } { \xi ^ { 3 } },
\end{eqnarray}
where $r _ { i k } = \left| \vec { r } _ { i } - \vec { r } _ { k } \right|$ represents the interatomic distance between atom $i$ and atom $k$, and $k _ { 0 } = \omega _ { 0 } / c$ with $\omega_0 = \left( \omega_j + \omega_{j^\prime} \right)/2$ and $\Gamma^{ii}_{jj'} = \gamma _ { j } =  2 \mu _ { j } ^2 {\omega_j}^3 / \left( 3 \epsilon_0 c^3 \right)$ is the spontaneous emission rate of a single atom on the $j$-th transition. The couplings for the energy shifts as well as the collective decays are plotted in Fig. \ref{couplings} as a function of the interatomic distance, whereas varying the dipole moment orientations leads to oscillations of various amplitudes (see Fig. \ref{geometry}). The terms $\Gamma^{12}_{12}$ and $\Omega^{12}_{12}$ are dipole-dipole cross coupling coefficients, which couple dipoles even though they are orthogonal.~\cite{agarwal2001vacuum}.

\section{Equilateral Triangle: Analytical Treatment}
For three 3-level atoms placed at the corners of an equilateral triangle with dipole orientations chosen such that the configuration features a $C_3$ symmetry (see Fig. \ref{geometry}), the states $\ket{\Psi_d^3}$ and $\ket{\Psi_{sr}^3}$ are both eigenstates of the Hamiltonian from eq. (\ref{Hamiltonian}) whose energies can be calculated explicitly. For three V-type atoms $\ket{\Psi_d^3}$ is given by
\begin{multline} \label{dark_state}
\ket{\Psi_d^3} = \frac {1}{\sqrt{6}} \{\ket{e_1 e_2 g} +\ket{g e_1 e_2}+\ket{e_2 g e_1}\\
	-\ket{e_1 g e_2}-\ket{e_2 e_1 g}-\ket{g e_2 e_1}\},
\end{multline}
whereas in the superradiant analogue $\ket{\Psi_{sr}^3}$, which is comprised of the exact same bare states, all signs are positive.
\begin{figure}[th]
	\centering
  \includegraphics[width=0.49\textwidth]{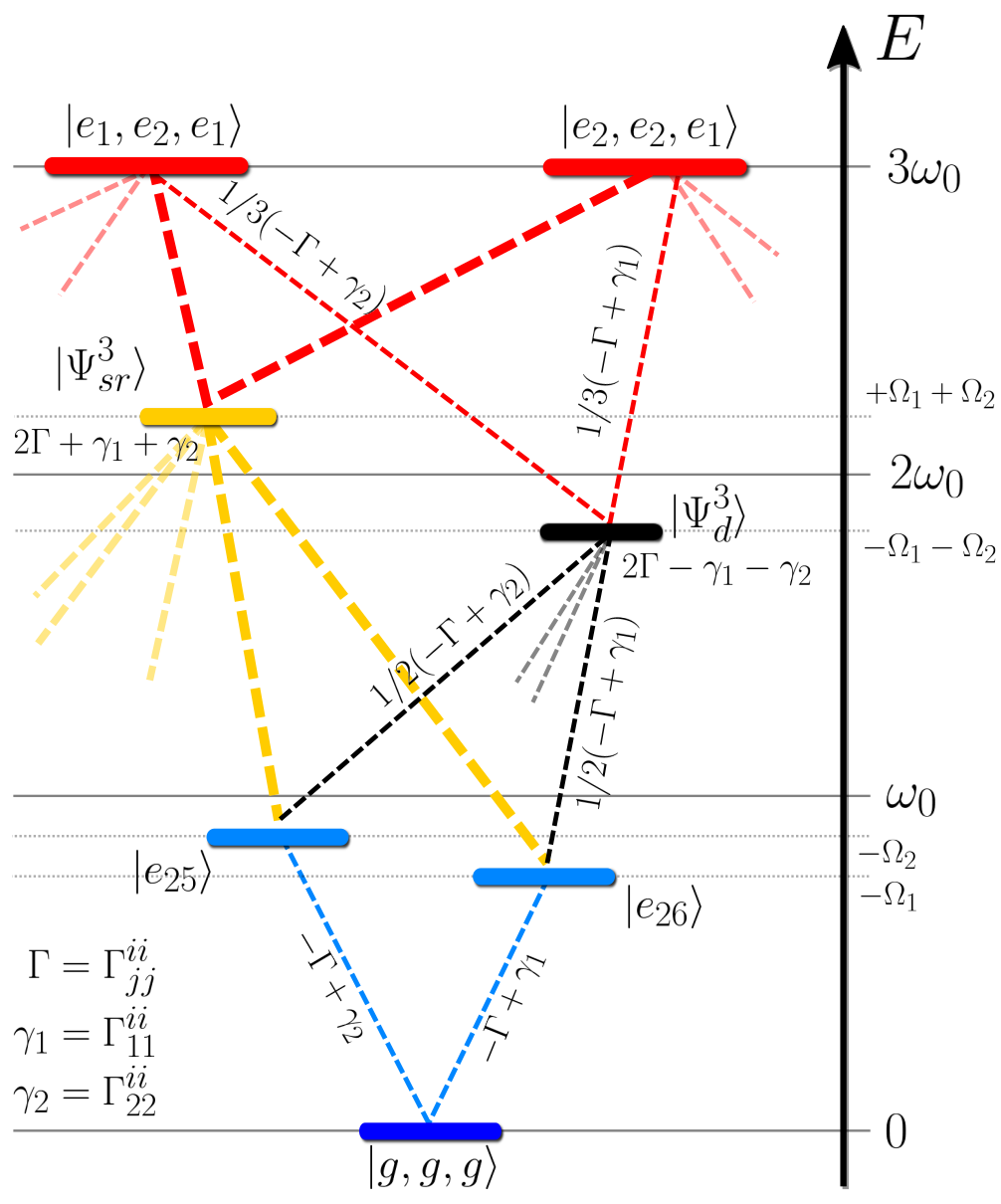}
	\caption{\textit{Decay Cascade.} After diagonalizing the Hamiltonian for the equilateral triangle configuration of three 3-level atoms with symmetric dipole orientations we show the decay cascade for selected eigenstates (the full cascade can be found in the supplement~\cite{SM}). We define $\gamma_1 \equiv \frac{3}{2} \Gamma P_I(k_0 r)$ and $\gamma_2 \equiv -\frac{3}{4} \Gamma P_I(k_0 r)+\frac{9}{8} \Gamma Q_I(k_0 r)$ and $\Gamma$ represents the spontaneous emission rate of a single V-type atom with degenerate excited states. Additionally, the collective energy shifts in the respective excitation manifolds are shown with $\Omega_k \equiv \Omega^{ii}_{kk}$ and $\omega_0$ being the energy between ground and excited states.}
	\label{cascade}
\end{figure}

Clearly, the dynamics of any eigenstate of the Hamiltonian is restricted to the decay towards other eigenstates $\ket{\psi_\mathrm{eig}}$ induced by the  Liouvillian, i.e.\ $\dot{ {\rho}}_\mathrm{eig} = \mathcal { L } \left[ { \rho }_\mathrm{eig} \right]$ with $\rho_\mathrm{eig} = \ket{\psi_\mathrm{eig}} \bra{\psi_\mathrm{eig}}$. The corresponding rates can be found by calculating the overlap with all other states. The decay and feeding rates for a certain selection of states are shown in Fig. \ref{cascade}. Explicitly, the decay rate for the eigenstate $\ket{\psi_\mathrm{eig}}$ is given by $\Braket{\psi_\mathrm{eig} | \mathcal { L } \left[ { \rho }_\mathrm{eig} \right] | \psi_\mathrm{eig}}$.

We find that the lowest lying energy state in the double excitation manifold corresponds to the antisymmetric dark state $\ket{\Psi_d^3}$, while the highest energy state is the superradiant state. With a more and more pronounced subradiance in $\ket{\Psi_d^3}$ at decreasing interatomic distances, also its feeding rate from higher lying states decreases, which culminates in a  decoupling from all other states and the electromagnetic field. In particular, for the equilateral triangle configuration, the lower an eigenstate lies energetically, the smaller its decay rate, as can be seen for selected states in Fig. \ref{cascade}. A full account of all coupling and feeding rates is available in the supplementary information~\cite{SM}. Also note that all feeding and decay rates to and from a particular state sum up to zero.

\section{Numerical Diagonalization for three and more Atoms}
For the case of $N \geq 3$ atoms we analyze the scaling of the decay rates as a function of the interatomic distance for increasing atom numbers and different geometries.

\begin{figure}[th]
	 \centering
	 \subfloat{\includegraphics[width=0.5\columnwidth]{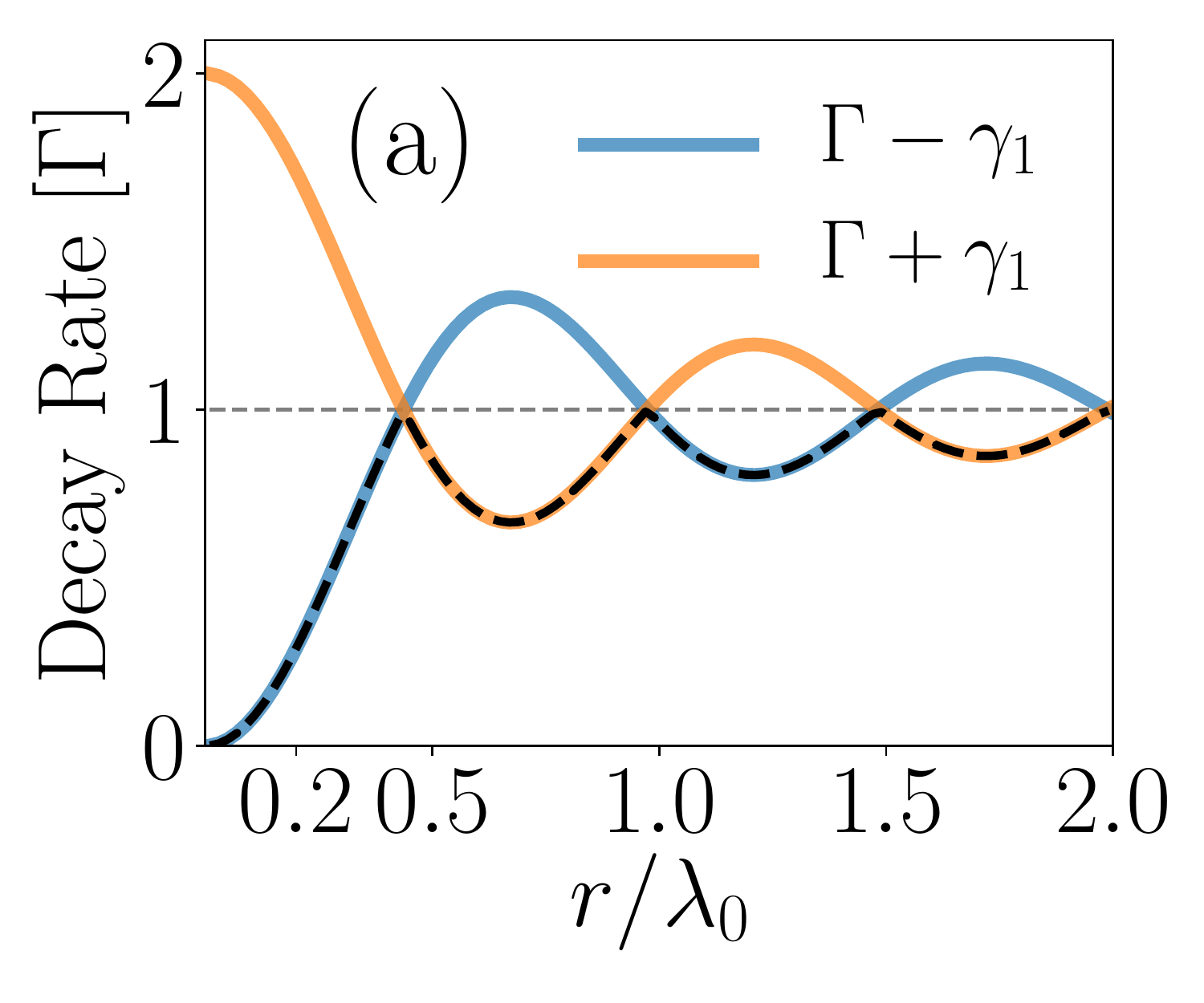}}
	 \subfloat{\includegraphics[width=0.5\columnwidth]{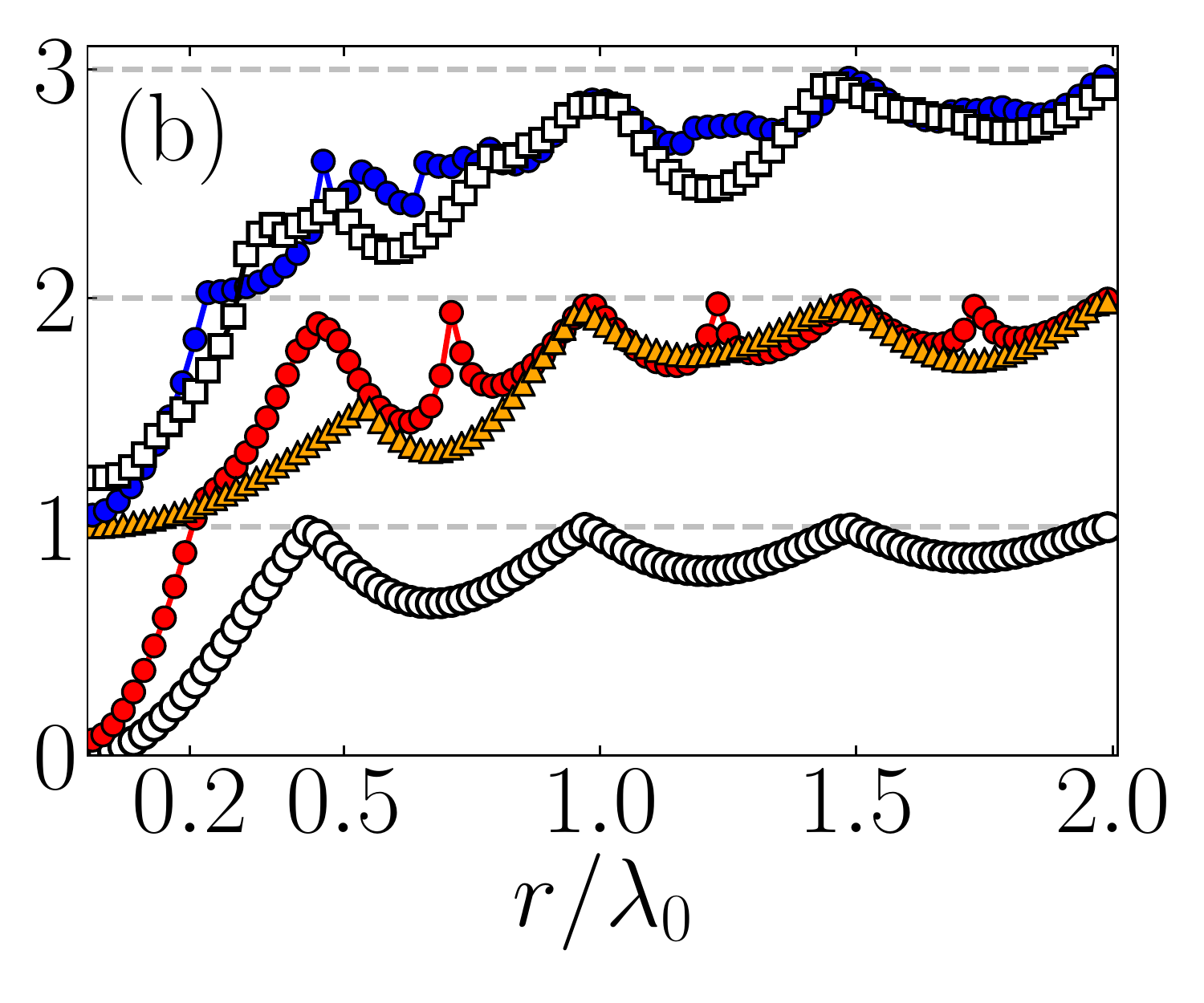}}
 		\caption{{\textit{Decay Rates.} (a) Sub- and superradiant decay rates for two dipoles as a function of the interatomic distance, where $\gamma_1 = \frac{3}{2}\Gamma P_I(k_0 r)$ and the black dashed line indicates the lowest decay rate at any given distance. (b) Lowest decay rates in the ($N-1)$ excitation manifold for $N$ $N$-level emitters as a function of the interatomic distance, with the white circles corresponding to two 2-level atoms, the red circles to three 3-level V-type atoms in a linear chain, the orange triangles to an equilateral triangle, the blue circles to four 4-level V-type atoms in a chain and the white squares to four atoms in a square.}}
	 \label{fig:rates}
\end{figure}

In Fig. \ref{fig:rates}(a) the simple case of two two-level atoms is shown, where the sub- and superradiant decay rates oscillate  around the independent decay rate $\Gamma$ with an amplitude decreasing with the interatomic distance, such that the super- and subradiant state switch their roles at each node. The black dashed curve corresponds to the lowest decay rate at any given distance and is generalized to more involved configurations in Fig. \ref{fig:rates}(b).

In Fig.~\ref{fig:rates} it can be seen, that the lowest collective decay rate for the $(N-1)$-excitation manifold for $N$ atoms goes to zero only if the interatomic distances approach zero, if all dipole transition moments are orthogonal to the plane of the atomic ensemble. For the equilateral triangle with symmetric dipole orientations and for $N\geq4$ atoms with more than two transitions this is not possible anymore and the minimal decay rate is $\Gamma$.

\section{Dark State Preparation}
\begin{figure}[th]
	 \centering
	 \includegraphics[width=0.4\textwidth]{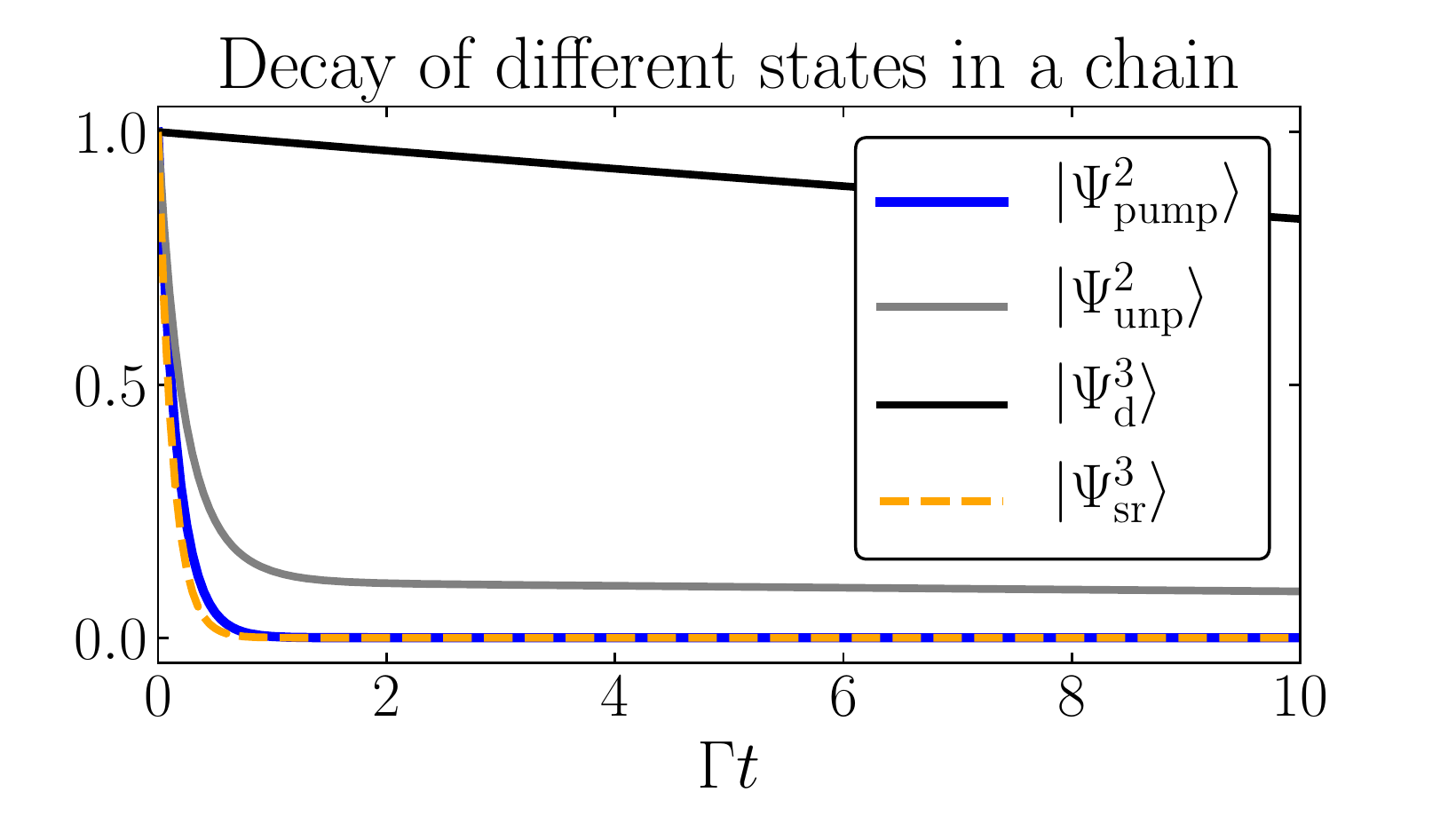}
	 \caption{\textit{Decay.} Population during decay for different doubly excited states in a linear chain of V-type atoms at $\lambda_0/50$ separation. The black line corresponds to the dark state $\ket{\Psi_{d}^3}$, the yellow dashed line to the superradiant state $\ket{\Psi_{sr}^3}$, the grey line to an unpolarized product state $1/{\sqrt{2}}(\ket{e_1 e_2}-\ket{e_2 e_1}) \otimes \ket{g}$ and the blue line to $1/{\sqrt{2}}(\ket{e_1 e_2}-\ket{e_2 e_1}) \otimes\ket{e_1}$, where all three atoms are initially excited.}
	 \label{fig:degeneracy}
\end{figure}

In most geometric configurations  apart from the equilateral triangle the anti-symmetric state $\ket{\psi_d^3}$ is not an exact eigenstate of the Hamiltonian from Eq. (\ref{Hamiltonian}) Yet, its subradiant property will prevail as shown for a linear chain in Fig.~\ref{fig:degeneracy}. The state $\ket{\Psi_\mathrm{unp}} = \left( \ket{e_1 e_2} + \ket{e_2 e_1} \right)/\sqrt{2} \otimes \ket{g}$ denotes a product state with atoms $1$ and $2$ entangled and exhibits subradiance as well. Generally, subradiance becomes particularly apparent at small atomic distances, where the derivative of the incoherent coupling with respect to $r$ is almost zero.

\begin{figure}[th]
	 \centering
	 \includegraphics[width=0.49\textwidth]{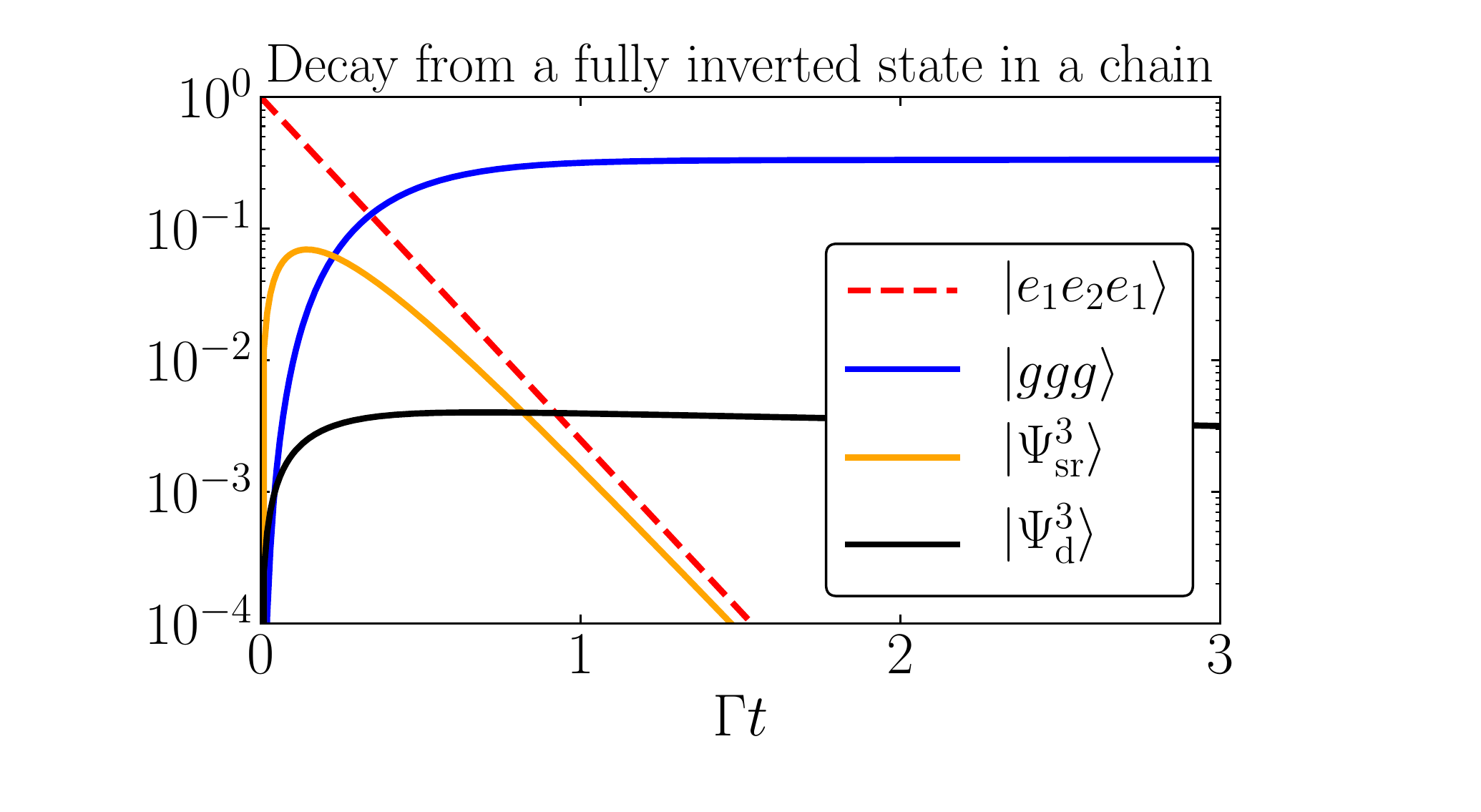}
\caption{{\textit{Inverted Decay.} Occupation probabilities during a purely dissipative preparation of different typical states of three V-type atoms in a linear chain with interatomic distance $\lambda_0/20$. The red dashed line represents the  inverted initial state, the orange line the superradiant state $\ket{\Psi_{sr}^3}$, the blue line the ground state fraction during the decay and the black line the dark state $\ket{\Psi_{d}^3}$ fraction during the decay process.}}
\label{fig:inverted_decay}
\end{figure}

At finite distances $\ket{\psi_d^3}$ can couple to to other states and will therefore decay as shown in \ref{fig:degeneracy}. Naturally, this means that it can be populated via decay from a higher lying state, which in this case are all triply excited states.  A typical case where the dark state becomes populated by photon emission for a three qutrit chain prepared in a totally inverted state, $\ket{e_1,e_2,e_1}$, is demonstrated  in Fig.~\ref{fig:inverted_decay}. Note that there are, in fact, eight different possibilities for triply excited states, i.e.\ $\ket{e_i,e_j,e_k}$ with $i,j,k \in \{1,2\}$, which lead to similar results. In Fig.~\ref{fig:inverted_decay} it can be seen, that the dark state can acquire a significant population, even via purely dissipative preparation, by choosing an appropriate geometric configuration. On the other hand, the feeding rate for the dark state becomes smaller with decreasing distances as it starts to decouple from the electromagnetic field.
\begin{figure}[th]
	 \centering
	 \subfloat{\includegraphics[width=.24\textwidth]{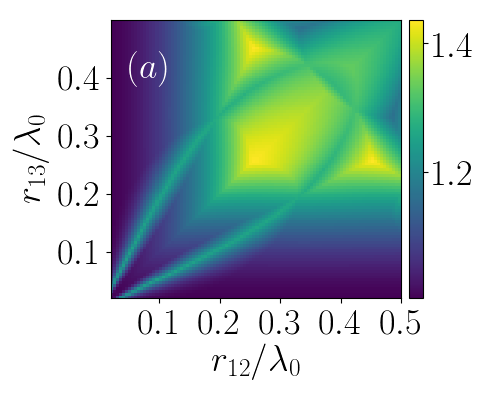}}
	 \subfloat{\includegraphics[width=.24\textwidth]{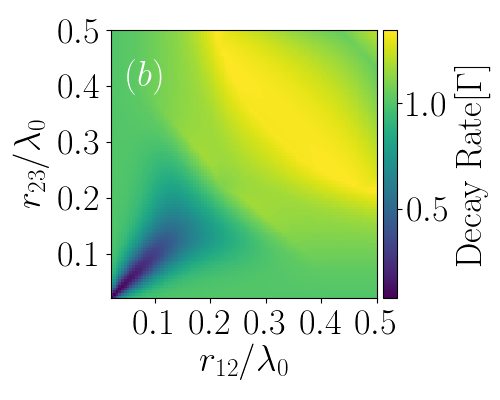} }
	 \caption{{\textit{Lowest Decay Rates.} The lowest decay rates (a) for three 3-level V-type atoms in a triangle configuration for eigenstates in the two-excitation manifold for different distances between atoms 1,2 and atoms 1,3 respectively and (b)  in a chain of atoms for different distances between atoms 1,2 and atoms 2,3 where all transitions are orthogonal to the direction of the chain are shown.}}
	 \label{lowestrate}
\end{figure}

As we have seen above, after an initial build-up of population in the dark state, the remainder of the population mostly ends up in the ground state. Hence, one can think of reusing the atoms in the ground state in order to further increase the occupation of the dark state. For this purpose , the preparation of the dark state $\ket{\Psi_d^3}$ or its superradiant analogue $\ket{\Psi_{sr}^3}$ can be facilitated by a continuous pump laser. It turns out that using different excitation phases for each atom can strongly improve the efficiency of this process, although this might be challenging to implement in practice.

We include a continuous pump in our model by adding the term ${H}_\mathrm{pump} = \sum_{i=1}^3 \sum_{j=1}^{2} \eta_i ({\sigma}^i_j+ {\sigma}^{i \dagger}_j)$ to the Hamiltonian with $\eta_1 = \eta$, $\eta_2 = \eta e^{i \varphi_1}$ and $\eta_3 = \eta e^{i \varphi_2}$, assuming that all atoms are driven with the same strength $\eta$. In our example the atoms are initialized in the ground state, $\ket{ggg}$, and we look at the population of the dark state after a given laser illumination time.

\begin{figure}[th]
	 \centering
	 \subfloat{\includegraphics[width=0.5\columnwidth]{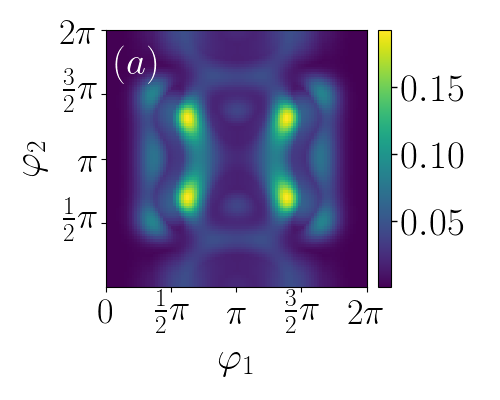}}
	 \subfloat{\includegraphics[width=0.5\columnwidth]{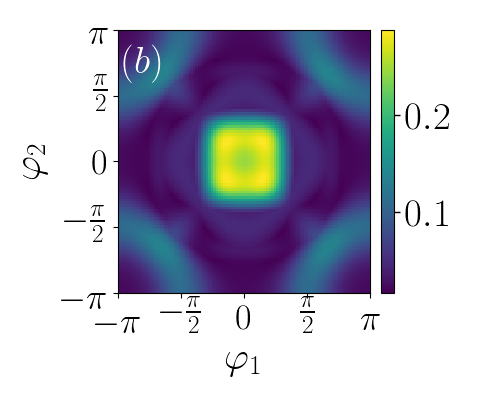}}
 		\caption{{\textit{Continous Laser Pump.} (a) Preparation probability of the dark state $\ket{\Psi_{d}^3}$ in a linear chain starting from $\ket{g,g,g}$ with interatomic separation $r = \lambda_0/50$, laser pumping strength $\eta = 8.5 \Gamma$ and laser phases $e^{i\varphi_1}$ and $e^{i\varphi_2}$ for atoms $2$ and $3$ on both transitions with respect to atom $1$ after $\Gamma t = 0.3$. (b) Probabilities for the superradiant state $\ket{\Psi_{sr}^3}$ in an equilateral triangle with the same parameters as in (a) where a maximum of $\approx 30 \%$ is obtained for phases $(\varphi_1,\varphi_2) = (\pm \pi/3, \pm \pi/3)$. }}
	 \label{fig:laser_preparation}
\end{figure}

In Fig.~\ref{fig:laser_preparation} the preparation probabilities for $\ket{\Psi_d^3}$ in a linear chain and for its superradiant analogue $\ket{\Psi_{sr}^3}$ in an equilateral triangle are shown as a function of the laser phase using a constant pump amplitude of $\eta = 8.5 \Gamma$ at an interatomic distance of $\lambda_0/50$ in both cases. For the linear chain it can be seen, that for instance if atom $2$ and $3$ are driven by phases $\varphi_1 = \pi/2$ and $\varphi_2 = \pi$ relative to atom $1$, the preparation probability for $\ket{\Psi_d^3}$ reaches $20 \%$. The state will still decay, but with a small rate, as given in Fig.~\ref{fig:degeneracy}. In contrast, setting $\varphi_1 = 0$ and $\varphi_2 = 0$ in the equilateral triangle results in a preparation probability of $30 \%$ for  $\ket{\Psi_{sr}^3}$. We find a surprisingly high preparation probability after a time evolution of $\Gamma t = 0.3$.

\begin{figure}[ht]
	 \centering
	 \subfloat{\includegraphics[width=0.5\columnwidth]{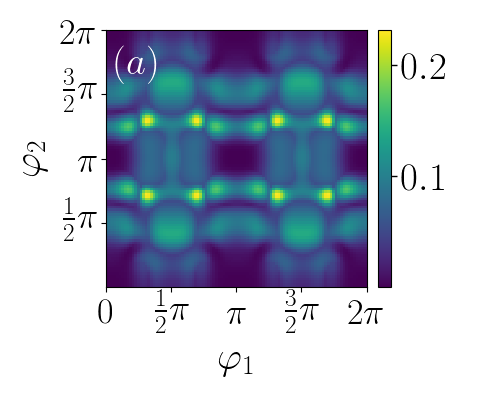}}
	 \subfloat{\includegraphics[width=0.5\columnwidth]{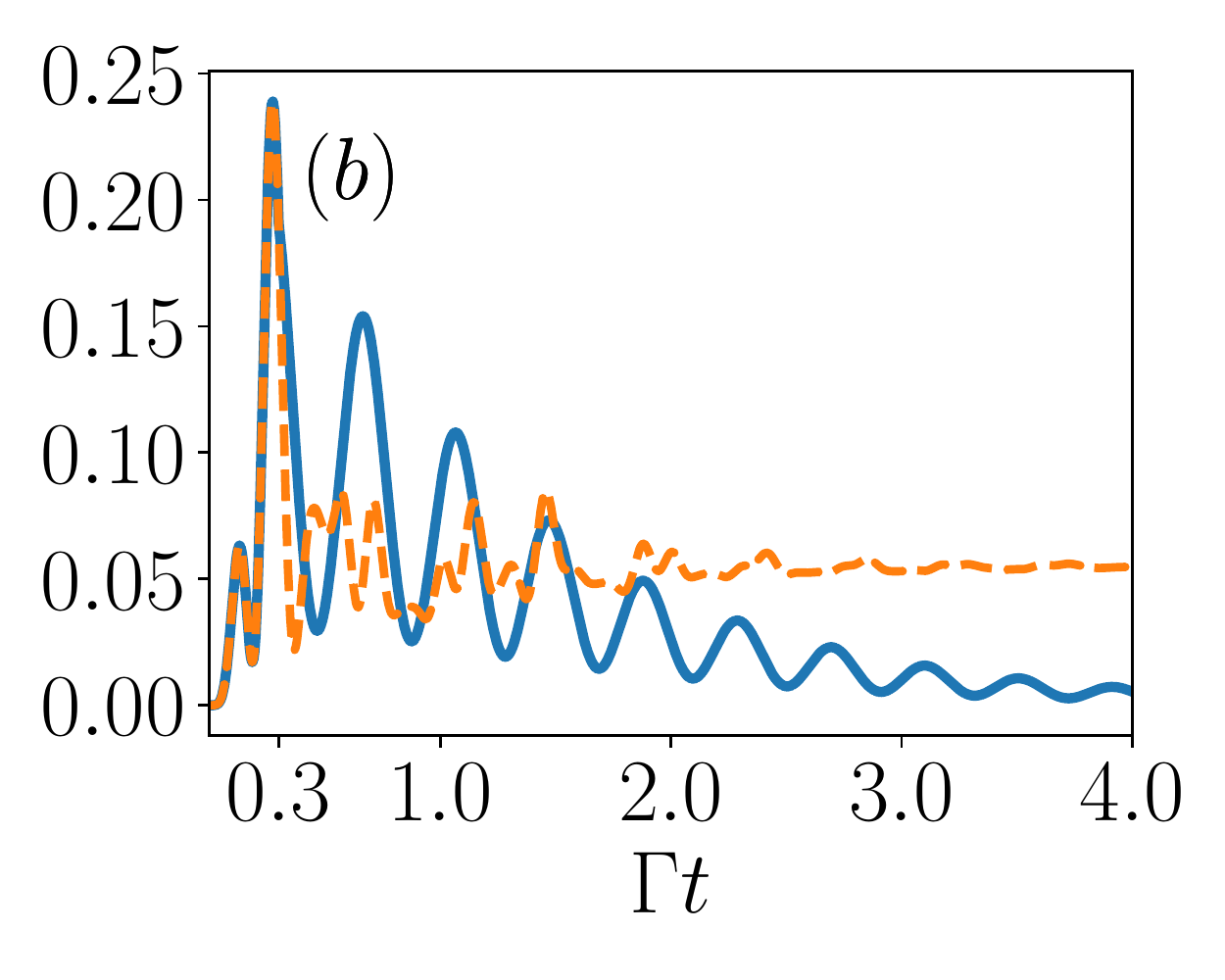}}
 		\caption{{\textit{Pulsed Laser.} (a) Preparation probability of the dark state $\ket{\Psi_{d}^3}$ in a linear chain starting from $\ket{g,g,g}$ with interatomic separation $r = \lambda_0/50$, laser pumping strength $\eta = 10 \Gamma$ and laser phases $\eta^1_j = \eta$,$\eta^2_1 = \eta e^{i\varphi_1}$,$\eta^3_1 = \eta e^{2i\varphi_1}$ and $\eta^3_2=\eta^2_2 =\eta e^{2i\varphi_2}$. (b) Probability for $\ket{\Psi_\mathrm{d}^3}$ with the parameters from (a) and phases $\varphi_1 = \varphi_2 = 7/10\pi$ where the preparation probability is maximal. The orange dashed line corresponds to continuous lasing throughout the time evolution and the blue line to a laser pulse for a time of $[\Gamma t = 0.03]$.}}
	 \label{fig:laser_preparation_phase}
\end{figure}

Now, we include different phases for different transitions by writing our pump Hamiltonian as ${H}_\mathrm{pump} = \sum_{i=1}^3 \sum_{j=1}^{2} \eta^i_j ({\sigma}^i_j+ {\sigma}^{i \dagger}_j)$. Figure~\ref{fig:laser_preparation_phase} (a) shows the preparation probability for $\ket{\Psi_{d}^3}$ for a range of different phases, where for instance for $\varphi_1=\varphi_2=7/10\pi$ a maximum of $\approx 24 \%$ is reached after $\Gamma t = 0.3$. In Fig.~\ref{fig:laser_preparation_phase} (b) we compare the time evolution of a pulsed laser with a continuous drive. Both cases lead to the same maximal value after $\Gamma t = 0.3$, but, after turning off the laser the dissipative dynamics lead to larger preparation probabilities shortly after that. Only for times longer than $\Gamma t = 1.5$ the laser driven system dominates the preparation probability. Specifically, for the case of pulsed lasing in Fig.~\ref{fig:laser_preparation_phase} (b) the first peak corresponds to a preparation probability of $24 \%$ and the second peak to $15 \%$, both within an evolution time of $\Gamma t = 1$.

\section{Conclusions}
We have generalized the concept of subradiance to multilevel emitters with several excited atomic levels decaying via independent decay channels towards a common ground state. In these systems the most subradiant states are completely anti-symmetric and maximally entangled. In contrast to ensembles of two-level emitters this multilevel type of dark states can hold several excitation quanta without decay. Hence detection could be facilitated by non-classical photon correlations at long time delays.

Entangled subradiant states have promising applications in quantum information processing and optical lattice clocks~\cite{optimal_geometry,ostermann2014protected}, amongst other key quantum technologies, where longer coherence times and a better understanding of energy level shifts induced via dipole-dipole interactions are crucial for improved accuracies.

States that do not decay as they decouple from the radiation field in turn are hard to access in order to prepare them directly. Yet, a probabilistic preparation can be achieved via spontaneous emission from higher lying states or in a much more efficient way by the application of laser pulses with spatial phase control.

Future work in this lines of studies will include coupling to a cavity field and analyzing the emission and absorption behaviour of multiple V-type emitters via an input/output formalism as in~\cite{input}. Another direction is the inclusion of vibrational degrees of freedom for each emitter as is demonstrated in~\cite{holstein,molecule} via a Holstein Hamiltonian for 2-level emitters.

\section*{Acknowledgements}

Financial support for this publication has been provided by the European Research Commission through the Quantum Flagship project iqClock (R.~H. and H.~R.) as well as by the Austrian Science Fund FWF through project P29318-N27 (L.~O.).

%

\pagebreak
\widetext
\begin{center}
\textbf{\large Supplemental Material}
\end{center}
\setcounter{equation}{0}
\setcounter{figure}{0}
\setcounter{table}{0}
\setcounter{page}{1}
\makeatletter
\renewcommand{\theequation}{S\arabic{equation}}
\renewcommand{\thefigure}{S\arabic{figure}}
\renewcommand{\bibnumfmt}[1]{[S#1]}
\renewcommand{\citenumfont}[1]{S#1}
\section{Decay Cascade For Three 3-level Emitters}
\begin{figure}[H]
	\centering
 	\includegraphics[width=0.99\columnwidth]{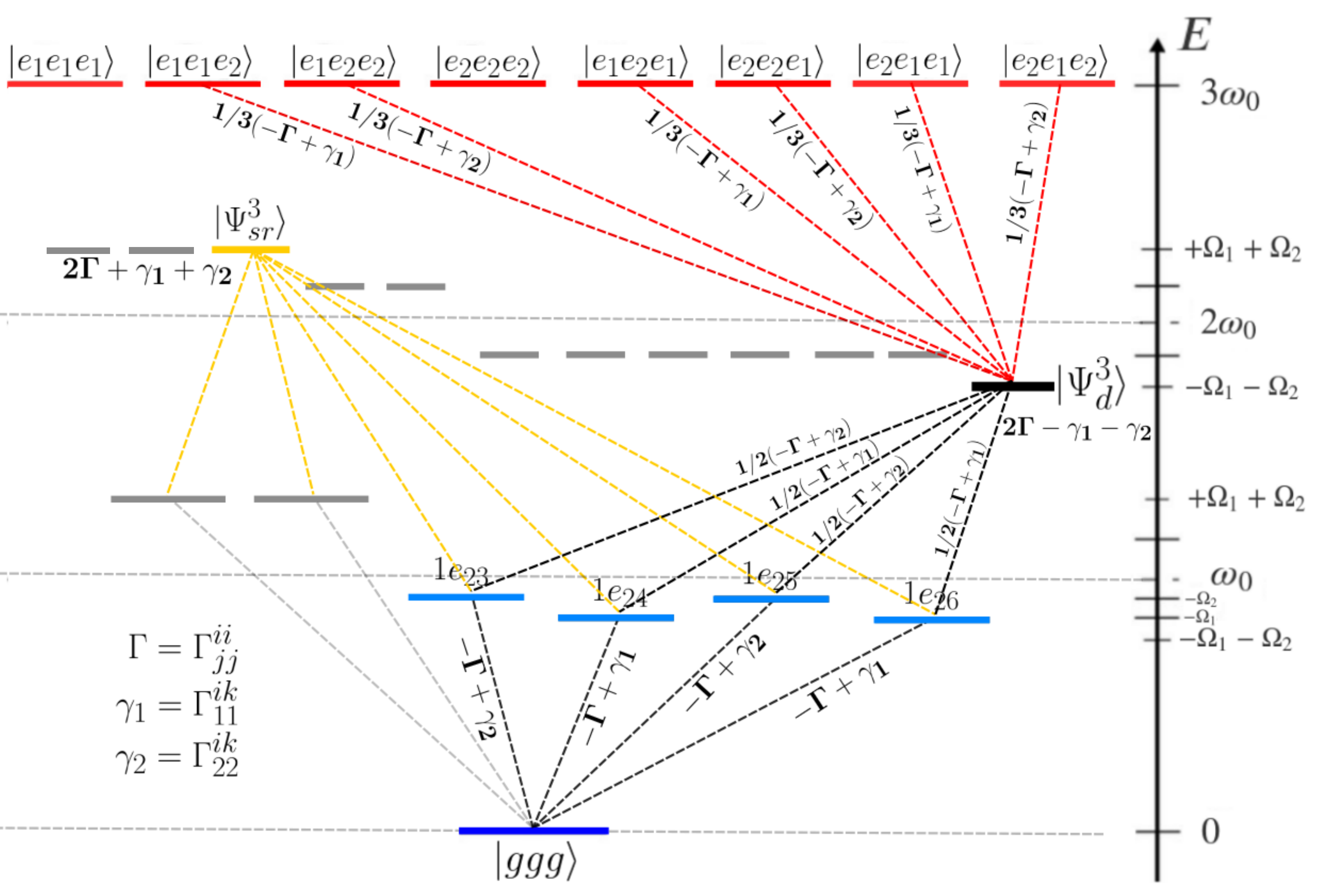}
	\caption{All $27$ eigenstates for the equilateral triangle with eight fully inverted states corresponding to all permutations of $\ket{e_i,e_j,e_k}$ with $i, j, k \in \lbrace 1, 2 \rbrace$, twelve eigenstates in the double excitation manifold, six eigenstates in the single excitation manifold and one ground state with no excitations. As in the main text, we define $\gamma_1 \equiv \frac{3}{2} \Gamma P_I(k_0 r)$ and $\gamma_2 \equiv -\frac{3}{4} \Gamma P_I(k_0 r)+\frac{9}{8} \Gamma Q_I(k_0 r)$ with $\Gamma$ being the spontaneous emission rate of a single 3-level V-type atom on both transitions, assuming degeneracy.}
	\label{cascade}
\end{figure}
By considering

\begin{equation}
\bra{\psi}\mathcal{L}[| \psi^3_d \rangle\bra{\psi^3_{d}}]\ket{\psi} \text{\quad and \quad} \bra{\psi}\mathcal{L}[| \psi^3_{sr}\rangle \bra{\psi^3_{sr}}]\ket{\psi}
\end{equation}
where $\ket{\psi}$ are all the lower lying eigenstates into which the superradiant and subradiant state decays, we obtain the decay rates into
the respective states. For $\ket{\psi} = \ket{\psi^3_{d}} $ or $\ket{\psi} = \ket{\psi^3_{sr}} $ we obtain the total decay rate.

Whereas for
\begin{equation}
\bra{\psi^3_{d}}\mathcal{L}[| e_i e_j e_k\rangle \bra{e_i e_j e_k}]\ket{\psi^3_{d}} \text{\quad and \quad} \bra{\psi^3_{sr}}\mathcal{L}[| e_i e_j e_k \rangle \bra{e_i e_j e_k}]\ket{\psi^3_{sr}},
\end{equation}
where $i,j,k = 1,2$ and $e_i e_j e_k$ are the 8 possible inverted states which can feed the Super- and Subradiant states, we obtain the feeding rates.
\begin{table}[H]
\begin{center}
    \begin{tabular}{| c | c | c | c |c | c | c | c |c |}
    \hline
    Feeding Rate from & $|e_1 e_1 e_1\rangle$ & $|e_1 e_1 e_2\rangle$ & $|e_1 e_2 e_2\rangle$ & $|e_2 e_2 e_2\rangle$ & $|e_1 e_2 e_1\rangle$ & $|e_2 e_2 e_1\rangle$ & $|e_2 e_1 e_1\rangle$ & $|e_2 e_1 e_2\rangle$ \\ \hline
    to $\ket{\psi^3_{d}}$ & $0$ & $-\frac{1}{3}\gamma+\frac{1}{3}\Gamma_1$ & $-\frac{1}{3}\gamma+\frac{1}{3}\Gamma_2$& 0 & $-\frac{1}{3}\gamma+\frac{1}{3}\Gamma_1$& $-\frac{1}{3}\gamma+\frac{1}{3}\Gamma_2$& $-\frac{1}{3}\gamma+\frac{1}{3}\Gamma_1$& $-\frac{1}{3}\gamma+\frac{1}{3}\Gamma_2$ \\ \hline
    to $\ket{\psi^3_{sr}}$ & $0$ & $-\frac{1}{3}\gamma-\frac{1}{3}\Gamma_1$ & $-\frac{1}{3}\gamma-\frac{1}{3}\Gamma_2$& 0 & $-\frac{1}{3}\gamma-\frac{1}{3}\Gamma_1$& $-\frac{1}{3}\gamma-\frac{1}{3}\Gamma_2$& $-\frac{1}{3}\gamma-\frac{1}{3}\Gamma_1$& $-\frac{1}{3}\gamma-\frac{1}{3}\Gamma_2$ \\ \hline
    \end{tabular}
    \end{center}
    \caption{Feeding rates where the negative sign means that it gives the decay rate for the energetically higher lying state.}
\end{table}
\begin{table}[H]
\begin{center}
    \begin{tabular}{| c | c | c | c |c | c | c | c |c |}
			\hline
			State & Total decay rate & 1 & 2 & 3 & 4 & 5 & 6 & $\ket{g,g,g}$ \\
		\hline
    $\ket{\psi^3_{d}}$ & $2\gamma -\Gamma_1-\Gamma_2$ & $0$ & $0$& $-\frac{1}{2}(\gamma-\Gamma_2)$ & $-\frac{1}{2}(\gamma-\Gamma_1)$ & $-\frac{1}{2}(\gamma-\Gamma_2)$ & $-\frac{1}{2}(\gamma-\Gamma_1)$ & $0$ \\ \hline \
    $\ket{\psi^3_{sr}}$ & $2\gamma +\Gamma_1+\Gamma_2$ & $-\frac{2}{3}\gamma-\frac{4}{3}\Gamma_1$ & $-\frac{2}{3}\gamma-\frac{4}{3}\Gamma_2$ & $-\frac{1}{6}\gamma+\frac{1}{6}\Gamma_2$ & $-\frac{1}{6}\gamma+\frac{1}{6}\Gamma_1$& $-\frac{1}{6}\gamma+\frac{1}{6}\Gamma_2$ & $-\frac{1}{6}\gamma+\frac{1}{6}\Gamma_1$ & $0$ \\
    \hline
    \end{tabular}
    \end{center}
\caption{Decay rates, where the first column is the total decay rate of the two states and the other columns are the decay rates to the lower lying eigentstates. The last column is the ground state and the other six are the six lower lying eigenstates in the one-excitation manifold.}
\end{table}

From the decay rate of the dark state we see that in the limiting case of infinitely close atoms, the decay could become even zero and the state would be indeed stationary under the Liouvillian superoperator. As is demonstrated in the main text, for the triangle it approaches $1\Gamma$ for infinitesimal distances and zero for the linear chain.

\end{document}